\documentclass{article}

\usepackage{arxiv}

\usepackage[utf8]{inputenc} 
\usepackage[T1]{fontenc}    
\usepackage{hyperref}       
\usepackage{url}            
\usepackage{booktabs}       
\usepackage{amsfonts}       
\usepackage{nicefrac}       
\usepackage{microtype}      
\usepackage{multirow}
\usepackage{multicol}
\usepackage{graphicx}
\usepackage[numbers,sort&compress]{natbib}
\usepackage{hyperref}
\usepackage{amsmath}
\usepackage{amssymb}
\usepackage{mathtools, nccmath}
\usepackage{float}
\DeclarePairedDelimiter{\nint}\lfloor\rceil
\usepackage[linesnumbered,ruled,vlined]{algorithm2e}

\title{Dealing with uncertainty in agent-based models for short-term predictions}

\author{
  Le-Minh Kieu\thanks{Corresponding author} \\
  School of Geography \& Leeds Institute of Data Analytics\\
  University of Leeds\\
  United Kingdom \\
  \texttt{m.l.kieu@leeds.ac.uk} \\
   \And
 Nicolas Malleson \\
  School of Geography\\
  University of Leeds and Alan Turing Institute\\
  United Kingdom \\
  \texttt{n.s.malleson@leeds.ac.uk} \\
  \And
 Alison Heppenstall \\
  School of Geography\\
  University of Leeds and Alan Turing Institute\\
  United Kingdom \\
  \texttt{a.j.heppenstall@leeds.ac.uk} \\
}

\begin{document}
\maketitle

\begin{abstract}
Agent-based models (ABM) are gaining traction as one of the most powerful modelling tools within the social sciences.
They are particularly suited to simulating complex systems. Despite many methodological advances within ABM, one of the major drawbacks is their inability to 
incorporate real-time data to make accurate short-term predictions. This paper presents an approach that allows ABMs to be dynamically optimised. Through a combination of parameter calibration and data assimilation (DA), the accuracy of model-based predictions using ABM in real time is increased.  We use the exemplar of a bus route system to explore these methods.  
The bus route ABMs developed in this research are examples of ABMs that can be dynamically optimised by a combination of parameter calibration and DA. The proposed model and framework can also be used in an passenger information system, or in an Intelligent Transport Systems to provide forecasts of bus locations and arrival times. 
\end{abstract}


\section{Introduction} 
\label{s:Intro}

Agent-based modelling (ABM) \citep{bonabeau_agent_2002} is a field that excels in its ability to simulate
complex systems. Instead of deriving aggregated equations of system
dynamics, ABM encapsulates system-wide characteristics from the
behaviours and interactions of individual agents e.g. human, animals
or vehicles. ABM has emerged as an important tool for many 
applications ranging from urban traffic simulation
\citep{balmer2009matsim}, humanitarian assistance
\citep{crooks_gis_2013} to emergency evacuations
\citep{schoenharl_design_2011}.

Despite the many advances and applications of ABM, the field suffers from a serious drawback: models are currently  unable to incorporate up-to-date data to make accurate real-time predictions \citep{lloyd_exploring_2016, wang_data_2015,
ward_dynamic_2016}. Models are typically calibrated once, using historical data, then projected forward in time to make a prediction. Here, calibration is ideal for one point in time, but as the simulation progresses, the prediction rapidly diverges from reality  due to underlying uncertainties \citep{ward_dynamic_2016}. These uncertainties come from \textit{dynamic} (changing over space and time), \textit{stochastic} (containing inherent randomness) and \textit{unobserved} (unseen from the data) conditions of the real system under study. An example of such a system can be found in bus routes. Each time a bus reaches a bus stop, the number of alighting passengers is uncertain and the number of waiting passengers downstream is unobserved. The bus route's conditions also change over time, e.g. traffic varies over the route and with at off-peak to peak periods. 
There are methods to incorporate streaming data into models,
such as \textit{data assimilation} (DA) routines
\citep{lewis_dynamic_2006, wang2000data}. Broadly, DA refers to a suite of techniques that allow
observational data to be incorporated into models
\citep{wang2000data} to provide an optimal estimate of the
evolving state of the system. Performing DA increases the
probability of having an accurate representation of the current state of
the system, thereby reducing the uncertainty of future predictions. This is a technique that has been widely applied
in fields such as meteorology, hydrology and oceanography \citep{kalnay_atmospheric_2003}.

There are, however, two methodological challenges that must be overcome to apply DA in ABM. First, DA methods are often intrinsic to their underlying models which are typically systems of partial differential equations with functions linearised mathematically.  Hence DA methods typically rely on linearising the underlying model \citep{harvey1990forecasting}. One of the most appealing aspects of agent-based models is that they are inherently non-linear, so it is not clear whether the assumptions of traditional DA methods will hold. Second, it is still unknown how much uncertainty DA can effectively deal with when implemented within ABM. Assimilation of real-time data into ABMs has only been attempted a few times and these examples are limited by their simplicity \citep{lloyd_exploring_2016, wang_data_2015,
ward_dynamic_2016}.

This paper is part of a wider programme of work\footnote{\url{http://dust.leeds.ac.uk/}} that is focused on developing DA methods to be readily used in ABM. This paper focuses on one particular model that aims  to make predictions of bus locations in real time. Bus route operation has been chosen due to its inherent uncertainties -- for example a model will need to account for uncertain factors affecting how buses travel on the roads \citep{khosravi2011prediction} -- but also for its tractability -- there are many fewer interactions than present in, say, a model of a crowd.  We also focus on one particular DA algorithm -- the Particle Filter (PF). This method is chosen due to its ability to incorporate data into non-linear models such as ABMs \citep{carpenter1999improved}.

The objectives of this paper are to: (1) perform dynamic state estimation to reduce the
uncertainty in the model's estimate of the \textit{current} system state; (2) improve the accuracy of short term forecasts.

All the numerical experiments in this paper will be tightly controlled, following an `identical twin' experimental framework \citep[for example see][]{wang_data_2015}. We will first develop a complex ABM of a bus route to generate fine-grained synthetic GPS data of buses, that are reasonably similar to real GPS data, for use as synthetic `ground truth' data. We call this model the `BusSim-truth' model. The next step
is to develop companion ABMs that are of simpler nature than BusSim-truth that will not know the parameters of BusSim-truth and will not have the dynamic and stochastic features of BusSim-truth. We will calibrate and evaluate these companion ABMs against the data generated from BusSim-truth. This experiment is designed to be similar to the real-time monitoring and predictions of bus locations, where models are often a simpler version of reality, that are calibrated to be as close as possible to reality. The prediction of bus location and arrival times are essential for bus operators and a topical research challenge \citep{bin2006bus}. The methods developed here can easily be applied to simulation and forecasting for \textit{real} bus systems and could, therefore, offer considerable potential impact. This is particularly pertinent in rapidly developing cities where daily bus schedules can be extremely erratic. In these cases accurate, up-to-date estimates of current waiting times will be highly beneficial to citizens who use (or would like to use) public transport. 

The contributions of this paper are threefold. First, several ABMs of bus routes are constructed that account for the interactions between the bus and passengers, the bus and the surrounding traffic, and between multiple buses are considered. While model development is not the sole focus of this paper, these bus route ABMs are novel and have utility for other studies. Second, this paper introduces a combination of parameter calibration and DA techniques that can dynamically optimise an ABM to enable accurate estimation of the bus system in real time. Third, this paper shows and quantifies the impacts of calibration and DA in dealing the with stochastic and dynamic nature of the system under study.

This paper is structured as follows. Section~\ref{s:problem} describes
the research problem and the related works in the literature.
Section~\ref{s:method} outlines the methodology. Section~\ref{s:experiments} describes
the numerical experiments that are conducted and discusses these
results. Finally, Section~\ref{s:conclusion} concludes the study and
considers the opportunities for future work. 

%
\section{Research problem and related works} \label{s:problem}

Historical and real-time bus GPS data is often used by operators to locate buses and predict their locations and arrival times. For instance, Public Transport Information and Priority System (PTIPS) is a state-of-the-art system in Australia to give priority to public transport vehicles on the roads and provide information on predicted bus arrival to passengers. The prediction of bus locations and arrival times in real time is a challenging problem \citep{chien2002dynamic}. Ideally, perfect 
knowledge of the current state of the system and any underlying processes is required.  However, obtaining this level of knowledge is impossible due to sources of uncertainty and the complex interactions in
bus operations. The majority of research within this area has focused on machine learning methods to find a direct mapping between input
data and bus arrival time. Examples of these methods include Artificial
Neural Networks \citep{chien2002dynamic}, Support Vector Machines
\citep{bin2006bus}, and Bayesian techniques
\citep{khosravi2011prediction}. While machine learning methods are generally
efficient in real time, they are solely reliant on the quality of available
data. Even with high-resolution datasets that record accurate spatio-temporal bus locations, the full complexity of the system will never be captured. 


There are analytical and simulation models of bus routes that aim to
reproduce the underlying processes in bus operations, and shed some
light on the associated uncertainties. One of the earliest successes in simulating a simple bus systems was from Cellular Automata modelling
\citep{luo2012realistic,o1998jamming,chowdhury2000steady,
	jiang2003realistic}. Whilst the dynamical foundations of these models
are well understood, they are outperformed by more sophisticated
models such as bus-following models
\citep{nagatani2000kinetic,huijberts2002analysis,Tang2012,
	nagatani2001bunching,hill2003numerical}; and traffic-following models
\citep{cats2010mesoscopic,toledo2010mesoscopic,hans2015real}.
Bus-following models aim to model the fundamental dynamics of a bus
route by modelling individual buses that follow each other (for example speeding up if the bus ahead is far away). Traffic-following models,
on the other hand, aim to model buses as a component of a transport
system with private and public transport, where their speeds are
affected by the traffic flow, traffic signals \citep{hans2015real} or
traffic density \citep{toledo2010mesoscopic}.

The majority of these models are \textit{static}, i.e. they only have parameters that are fixed over time. We can represent these static simulators 
with the equation $Y = f(X)$, where $f$ represents the simulator. A run of a simulation is defined
as the process of producing one set of data $Y$ for a single set of
model parameters $X$. One way for these
models to reduce their uncertainty and fit more closely to the observed data is to
adjust the model parameters until the model satisfies some predetermined
criteria. This parameter adjustment process is often referred to as
\textit{parameter calibration}. Popular optimisation techniques include simulated annealing~\citep{pennisi_optimal_2008}, genetic
algorithms,~\citep{heppenstall_genetic_2007, malleson_optimising_2014},
and approximate Bayesian computation~\citep{grazzini_bayesian_2017}.
Parameter calibration, especially with ABMs, is often only
implemented once, and therefore cannot account for any changes that may take place within
the system. In the traffic context these might include accidents, traffic signal failures, vehicle faults, etc. Static models are simple to implement, but struggle to model dynamic systems.


In real-time applications, e.g bus location or bus arrival time prediction in real time, prediction models often have to deal with the fact that there are so much uncertainty in bus operations. Real bus operation is \textit{dynamic} (changing over time) and also \textit{stochastic} (contains inherent randomness). In real time, there are also many unobservable information of bus operation, such as the number of passengers who are waiting at
downstream stops or the number who plan to get off the bus, and the
surrounding traffic conditions.  The lack of information about these factors means that any model of bus
operation in real time will have to make assumptions thereby introducing uncertainties. 

Therefore, this research will explore a combination of parameter calibration and a \textit{data
	assimilation} (DA) technique to calibrate a dynamic ABM bus route simulator using historical data, and then dynamically optimised it on-the-fly using real-time data. This, in itself, is a novel and important
contribution. Few previous efforts have attempted to incorporate data
assimilation with agent-based models \citep[for example
see][]{ward_dynamic_2016, wang_data_2015}, and it is unclear how DA
methods, that have typically been created for linear models
\citep{harvey1990forecasting}, can be adapted for non-linear ABMs. 

DA methods assume that observational data are sparse and only describe
the target system in limited detail. Therefore a model is essential as a
means of filling in the gaps in space and time left by the observations through the generation of additional data.
In effect, the model propagates data from observed to unobserved areas
\citep{carrassi_data_2018}.  Although techniques can be used to
perform parameter estimation, they are most often framed as a state
estimation problem. The aim is to calculate a posterior probability for
the state vector $X_t$, given prior distributions from a model (in this
case, a bus route operation model) and data from observations. It is
this marriage of a model and real-time observational data (and the 
associated uncertainties) that offers the means of allowing all the
available information to be used to determine the true state of the
system as accurately as possible \citep{talagrand_use_1991}.

Models where the system state at time $t$ are only dependent on the state at time $t-1$ are termed \textit{Markovian}. We are particularly interested in ABMs that can be written in a Markovian nature because DA algorithms require knowledge to the \textit{full} model state in the form of the state vector $X_t$. While some ABMs in the literature track agent histories and use this information to decide future states, these can be recast as Markovian ABMs by expanding the state vector to include these histories. Implementing a bus route system as a Markovian model requires variables such as vehicle locations, speeds, occupancies etc. It is reasonable to
assume that the system state at the next time step only depends on the
value of these variables at the current time step. For simplicity, we
assume that the state vector used here has a fixed size.  The unused variables (i.e. those for buses that have yet to enter the system) can be set to zero, enabling the state vector to be treated as sparse and
passed efficiently between iterations. If the state vector has a fixed
size, then all possible states of the system belongs to a state-space
$\mathcal{X} \in \mathbb{R}^n$. The system state evolves in some fixed
interval \{0,..,K\}. We denote the state of the bus route at time $t$ by
$X_t \in \mathcal{X}$.

\begin{figure}[H]  \centering\includegraphics[width=13cm]{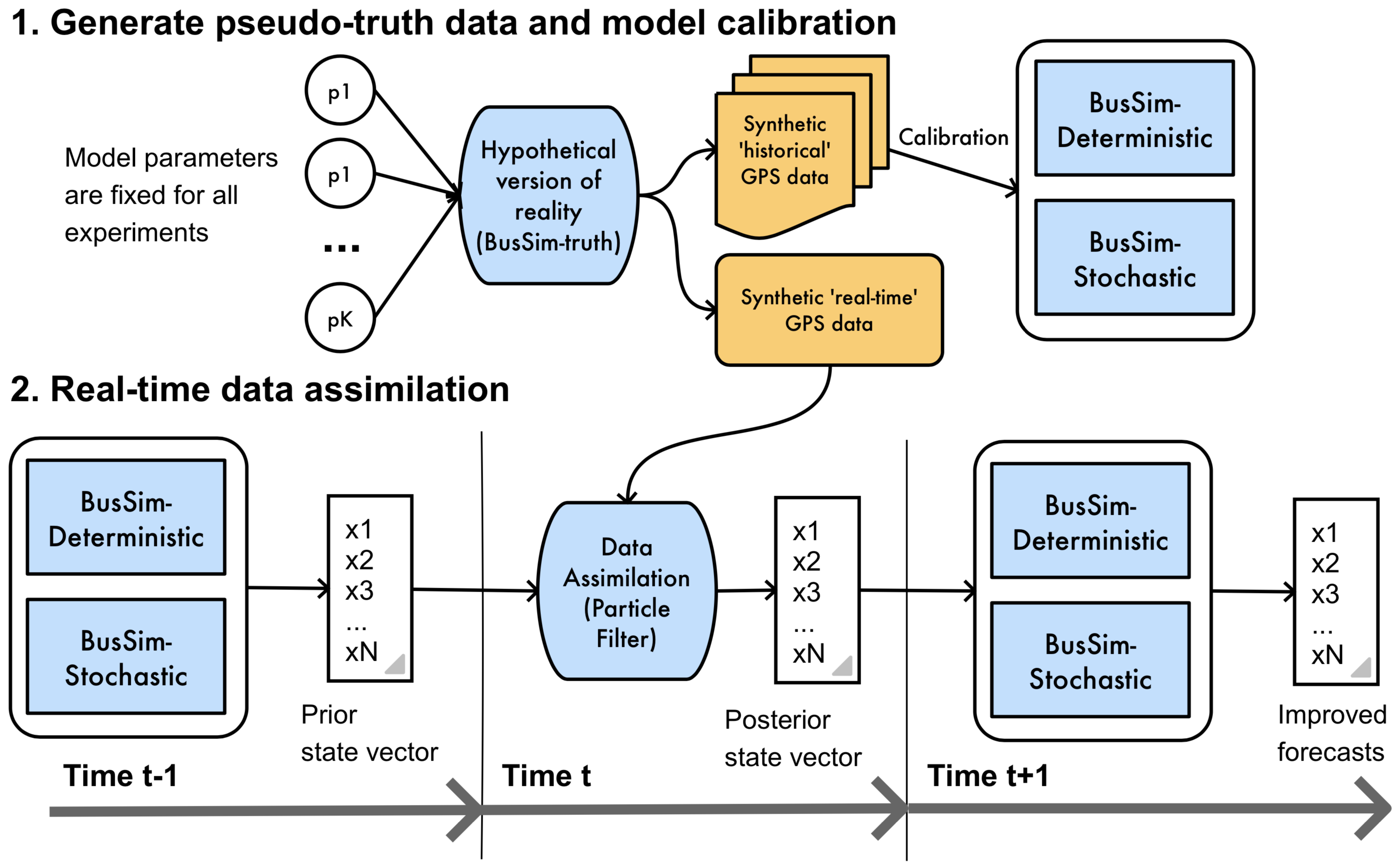}
	\caption{Study workflow.} 
	\label{fig:workflow} 
\end{figure}

This paper follows an `identical twin' experiment framework \cite{wang_data_2015}, where experiment data to be used will be generated from simulation, instead of using real data. The reason is that real data often comes with noise that hides the true state of the bus route (e.g. noise from GPS data). A simulated synthetic data would enable us to control the level of noise in the data, and to evaluate the modelling results against the ground truth rather than noisy data. Figure \ref{fig:workflow} shows the workflow of this study. 

The study workflow generally consists of 2 major steps. It starts with the development of a Markovian ABM of bus route operation that will be referred as \textit{BusSim-truth}. BusSim-truth is a hypothetical version of reality and will be used to generate synthetic GPS data of bus locations with timestamps. Two sets of data will be generated. The first represents `historical' GPS data, which are essentially the outputs of multiple runs of the same BusSim-truth model with the same predefined set of parameters. The GPS data will be slightly different each time the model is run because BusSim-truth is stochastic (its outputs vary slightly from one run to another) and dynamic (the parameters that control factors such as the amount of traffic vary during a single model run). The second set of data represent a single run of BusSim-truth, also using the same set of parameters. These data will represent synthetic `real-time' GPS data and will be used to conduct data assimilation. This situation is similar to the reality, where `historical' data across multiple days are used to calibrate models and `real-time' data represent the \textit{current} state of the world. BusSim-truth will be reasonably realistic and will replicate popular phenomenon in bus operations such as bus bunching (two buses of the same line arrive at the same bus stop at the same time). 

In reality, any simulation model is a simplification of the actual dynamics. Taking this into consideration, we develop two simpler variations of BusSim-truth, knowing that they would not be able to perfectly represent the dynamics in BusSim-truth.  The two variations are: 
\begin{itemize} 
	\item \textbf{BusSim-deterministic}. This model evolves exactly the same way in each model run;
	\item \textbf{BusSim-stochastic}. This model is stochastic, e.g. the numbers of people waiting at bus stops is drawn from a random distribution 
\end{itemize} 

As would be necessary in reality, BusSim-deterministic and BusSim-stochastic will first be calibrated against the synthetic `historical' GPS data. In the second step of the study workflow, DA will be used in an attempt to update the states of the models to the `real-time' GPS observations in order to produce more accurate short-term forecasts of the system behaviour. 

\section{Methodology\label{s:method}}



\subsection{A hypothetical version of reality: BusSim-truth and its two simpler variations}

The first step in the proposed workflow is to develop an agent-based bus route model that will be used to generate synthetic GPS data for each bus on the route (BusSim-truth). BusSim-truth is a stochastic and dynamic model with two classes of agents (bus and bus stop) and predefined parameters (see Table \ref{tab:Agents} ). It is stochastic because the number of boarding passengers is drawn from a random distribution, and dynamic because it parameters gradually change over time. The level of stochasticity and dynamicity in BusSim-truth can also be adjusted to represent bus route systems where conditions are largely stable or volatile over time.  

\begin{figure}[ht]
	\centering
	\includegraphics[width=1\textwidth]{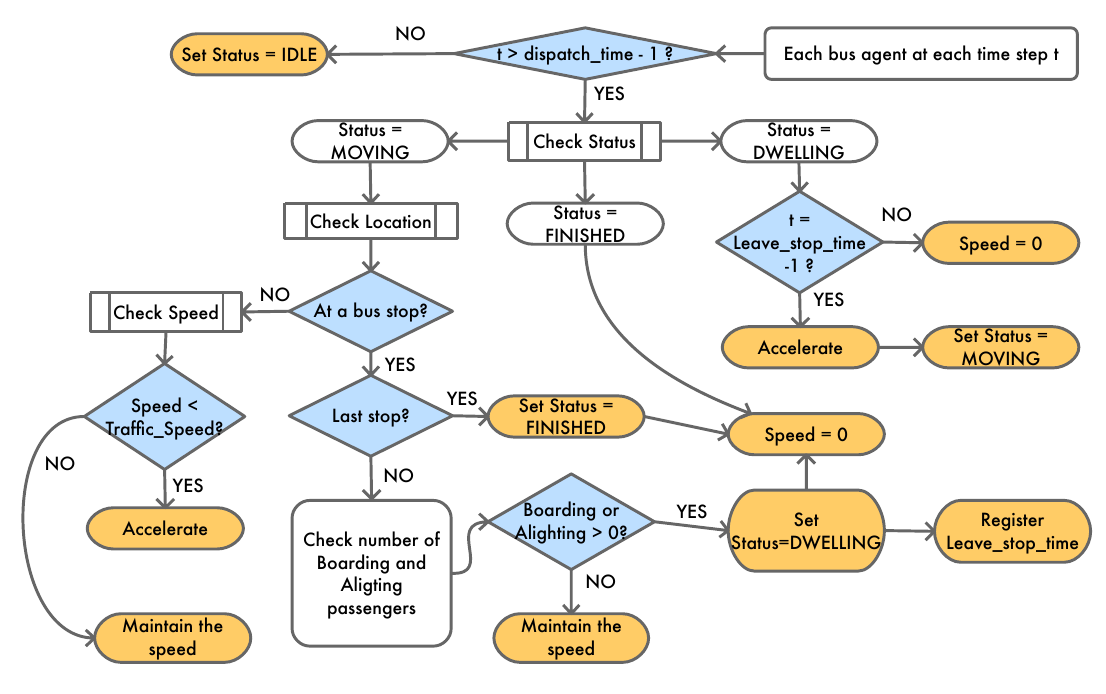}
	\caption{Flowchart of BusSim-truth.}
	\label{fig:BusSim_flowchart}
\end{figure}

Figure \ref{fig:BusSim_flowchart} illustrates the workflow for BusSim-truth. Only a brief explanation of the model is included here, as more information on how the BusSim-truth model works can be found in the Appendix A. 
At each current time step, each Bus agent checks whether the next time step would be larger than the vehicle's scheduled dispatch time. If it is, we then check whether the bus is on the road (Status equals $MOVING$), or at a stop for passenger dwelling (Status equals $DWELLING$), or has finished its service (Status equals $FINISHED$), otherwise the bus remains $IDLE$.  

If the status is $MOVING$, we first check whether the bus is at a bus stop, by comparing the $GeoFence$ area of each bus stop agent with the bus' location. If the bus is not approaching a bus stop, its current speed will be compared with the surrounding traffic speed. In the case it is slower, we assume that the bus will speed up. If the speed already matches the traffic speed, the bus will maintain the same speed. Currently the traffic volume on the whole network is represented as a single dynamic parameter, although in practice it would be relatively trivial to make the traffic volume heterogeneous across the network. The system will first check if the stop is at the last stop when the bus is approaching a bus stop, where the bus' status will be changed to $FINISHED$ and the bus speed changed to zero. If it is not the last stop, the system will change the status of agent Bus to $DWELLING$ and its speed to zero. 

\begin{table}[htbp]
	\centering
	\caption{Type of agents and their parameters in BusSim-truth}
	\resizebox{\textwidth}{!}{\begin{tabular}{ll}
			\textbf{Parameter} & \textbf{Description} \\
			\midrule
			BusID  & Unique ID of the bus agent \\
			Acceleration & The acceleration value in $m/s^2$ if the bus needs to accelerate \\
			StoppingTime & Deadtime due to door opening and closing if the bus has to stop \\
			Visited & List of visited bus stops \\
			States & Whether the bus is idle, moving, dwelling or finished \\
			Trajectory & GPS coordinates of bus locations \\
			\midrule
			bus stopID & Unique ID of the bus stop \\
			Position & Distance from the first stop \\
			$Arr_m$ & Passengers arrived to the stop per second \\
			$Dep_m$ & Percentage of onboard passengers alight at the stop \\
			Arrival\_time & Store actual arrival time of buses at the stop \\
			GeoFence & A circle area to identify whether the bus is at the bus stop \\
			\bottomrule
	\end{tabular}}%
	\label{tab:Agents}%
\end{table}%

\begin{figure}[h]
	\centering
	\includegraphics[width=1\textwidth]{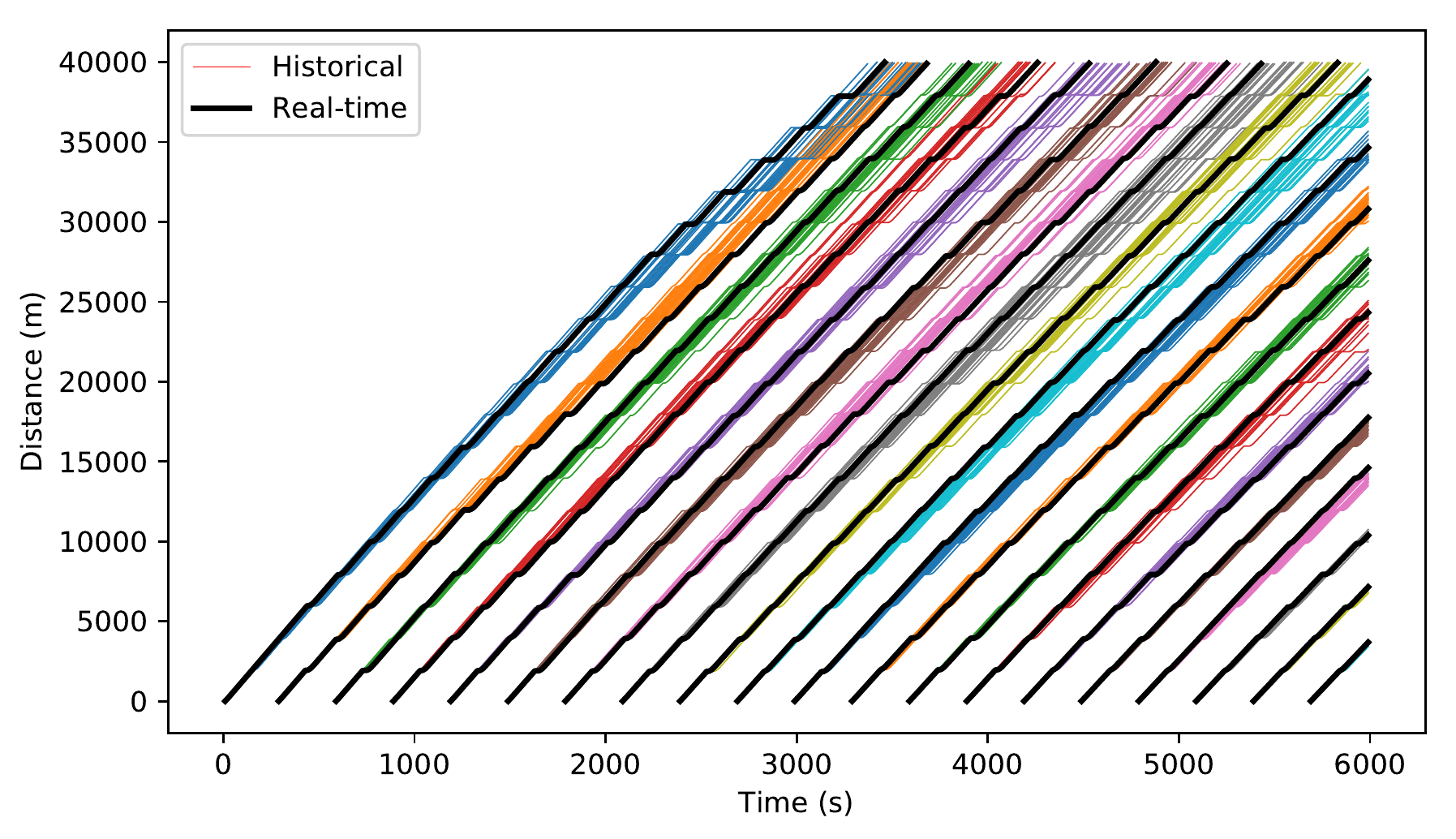}
	\caption{Synthetic `historical' versus `real-time' GPS bus location data. Each coloured line shows the trajectory of one bus in the `historical' GPS data. As the BusSim-truth model is stochastic, there are differences between the trajectories. This is similar to the reality where buses operate slightly differently on multiple days. The bold black lines are another instance of bus trajectory that we consider as the `real-time' GPS data.  }
	\label{fig:historical_realtime}
\end{figure}

As described in Section \ref{s:problem}, we use BusSim-truth to generate two sets of synthetic data: (1) `historical' GPS data that simulate normal bus route operation over a number of days and are used for calibration; and (2) `real-time' GPS data that represents a single run of the model and are used to represent the bus system \textit{today}. These are visualised in Figure~\ref{fig:historical_realtime}. Each record in the synthetic data is called an \textit{observation vector}. The vector contains all of the observations made from the `real world' (in this case the BusSim-truth).

\subsection{Optimising the parameters of BusSim-deterministic and BusSim-stochastic}
\label{s:calibration}

Most agent-based models have a large number of parameters. For the BusSim models, the model parameter vector $S_t$ at time $t$ contains the arrival rate $Arr_m^t$, departure rate $Dep_m^t$ at each stop $m$, and the traffic speed $V^t$. 
\begin{equation}
S_t  = \left[ \begin{array}{ccc}
Arr_m^t & Dep_m^t & V^t  
\end{array} \right] \quad m=1..M 
\label{eq:S}
\end{equation} 

The two simpler models (BusSim-deterministic and BusSim-stochastic) first need to be calibrated to the observations (the historical data). Here an automatic parameter calibration process, based on the Cross-Entropy Method (CEM) \citep{rubinstein1999cross} is used. CEM is a population-based Monte Carlo learning algorithm to combinatorial multi-extremal optimisation and importance sampling. It originated from the field of rare event simulation, where even small probabilities need to be estimated \citep{rubinstein1999cross}. In principle, CEM develops a \textit{probability distribution} over possible solutions for the optimal parameters of the model. New solution candidates are drawn from this distribution and are evaluated. The best candidates are then selected to form a new improved probability distribution of the optimal parameters, until certain criteria are met. CEM is chosen over some other popular optimisation methods in parameter calibration of ABMs, such as Genetic Algorithm \citep{heppenstall_genetic_2007} and simulated annealing~\citep{pennisi_optimal_2008}, because of its probabilistic nature that facilitates the calibration of stochastic models \citep{ngoduy2012calibration}. The interested reader may refer to \citep{rubinstein1999cross}, and various applications of CEM, such as \citep{ngoduy2012calibration}, for a more detailed account. A pseudo-code of the CEM algorithm that we adopted for this paper has also been described in Appendix B. 

Formally, the parameter calibration is an optimisation problem to minimise some performance index $PI(\pi)$ over all $\pi \in \mathbb{R}^k$. Here a solution $\pi= (\pi_1,\pi_2,...,\pi_k)$ denotes a set of parameters of the model under consideration and $k$ denotes the number of dimension in this set. Let $\pi_*$ denote the optimal solution, or the best set of model parameters that we want to find, that is:

\begin{equation}
\pi_* = \text{argmin} \quad PI(\pi), \quad \pi \in \mathbb{R}^n
\end{equation}

The above objective function is equivalent to finding $\pi_*$ such that $PI(\pi_*) \leq PI(\pi) \quad \forall X \in \Pi$, where $\Pi$ is a constrained parameter space such that $\Pi \in \mathbb{R}^k$. The performance index $PI(\pi)$ is generally the difference between model output and observed data. The complexity of this problem comes from the stochasticity of BusSim, where the same solution $\pi$ may yield a different realisation $PI(\pi)$. To reduce this stochastic effect, it is necessary to run the (stochastic) model multiple times, and to evaluate the simulation outputs against a compilation of observed data from multiple days or instances. Let $K_I$ be the number of replications required for each model evaluation and $K_O$ be the number of instances in the observed data, we can derive a more detailed objective function of the parameter calibration problem: 

\begin{align}
\text{min} \ PI(\pi) = \frac{1}{N \cdot T} \sum_{t=1}^T \sum_{n=1}^N \Bigg[ \bigg|  \frac{1}{K_I} \sum^{K_I}  s_{j,i,t}^{SIM}  - \frac{1}{K_O} \sum^{K_O}  s_{j,o,t}^{OBS} \bigg| + \nonumber \\
\bigg| \sqrt{\frac{\sum^{K_I} \big( s_{j,i,t}^{SIM} - \hat{s}_{j,i,t}^{SIM} \big)^2}{K_I-1}}  - \sqrt{\frac{\sum^{K_O} \big( s_{j,o,t}^{OBS} - \hat{s}_{j,o,t}^{OBS} \big)^2}{K_O-1}} \bigg|   \Bigg]
\label{eq:objective_function}
\end{align}

Where $N$ is the number of buses, $T$ is the number of time steps, $s_{j,i,t}^{SIM}$ is the location of simulated bus agent $j$ at time $t$ for the replication $i$, and similarly $s_{j,o,t}^{OBS}$ is the synthetic observed location of bus $j$ at time $t$ for the instance $o$. The objective function in Equation \ref{eq:objective_function} can be seen as the sum of the difference in mean location and standard deviation of locations at each time step for each bus and each replication/instance between simulated outputs and synthetic observed data. We want to evaluate the difference in not just the mean but also the standard deviation of bus locations because the system under study is stochastic, so it is not just the mean but also the spread of bus locations over multiple instances are important. 

\subsection{Data Assimilation using a Particle Filter (PF)}
\label{s:PF}

We can formulate an ABM as a state-space model $\dot{X}_t = f(X_{t}) + \epsilon_t$ and use data assimilation (DA) to dynamically optimise the model variables with up-to-date data to reduce uncertainty. The state-space model is represented by a state-space vector $X_t$ at time $t$, which contains all information of the current state of each agent in the model: 
\begin{align}
X_t = \left[ O_t \quad S_t \right] \nonumber \\
=\left[ \begin{array}{ccccccc}
c_j^t & s_j^t & v_j^t & Occ_j^t & Arr_m^t & Dep_m^t & V^t \end{array} \right] 
\label{eq:state_vector}
\end{align}

The state-space vector $X_t$ must contain all of the information that identifies the current state of the modelled system, allowing it to be projected forward to the next time step. Thus vector $X_t$ usually contains both the observation vector $O_t$ and the model parameters vector $S_t$ (Equation \ref{eq:S}) at time $t$. Note that $S_t$ has been calibrated in the previous section, but is still included in the state space vector $X_t$ to allow the model to be dynamically optimised with new data -- this is essential in dynamic situations where parameter values change over time. This approach is often referred to as dynamic calibration \citep{eicker2014calibration}.

Data Assimilation (DA) is a suite of methods to adjust the state of a running model using new data to better represent the current state of the system under study \citep{ward_dynamic_2016}. DA was born out of data scarcity, where observation data are sparse and insufficient to describe the system. Notwithstanding the proliferation of new data sources, insufficient data is still a major problem in research. The prediction of bus locations is a clear example where the number of future boarding and alighting passengers are unknown in real time. DA algorithms fill in the spatio-temporal gaps in the observed data by running a model forward in time until new observed data are available. This is typically called the \textit{predict} step in DA algorithms. After the predict step, DA has an estimate of the current system state and its uncertainty (which is often referred as the `prior' in Bayesian literature). The next step is typically called the \textit{update} step, where new observations and uncertainty are used to update the current state estimates. The result is referred to as the `posterior' in Bayesian literature, and should be the best guest of the system state from both the observations and model.

There are several DA algorithms in the literature, ranging from the simple Kalman Filter \citep{meinhold1983understanding} to more advanced extensions, including extended, ensemble and unscented Kalman Filter \citep{ward_dynamic_2016}. These algorithms generally aim to extend the original Kalman Filter by relaxing the assumption of linearity and introducing methods to work with non-linear models. However, they may not be the most suitable candidate to incorporate data into ABMs for two reasons. First, ABMs are driven by a large number of interacting agents with goals, history and behavioural rules. As a result, they lack an analytic structure, such as differential or difference equations, to facilitate the implementation of the Kalman Filter and its extensions where often the model Jacobian and covariance matrices need to be formulated \citep{wang_data_2015}. Second, although the assumption of linearity has been relaxed, these extensions assume that the noise in the model estimation is Gaussian. 

There is a flexible Bayesian filtering method that has been designed to work with non-linear, non-Gaussian models without analytical structure; this is the Particle Filter (PF). The key idea is to approximate a posterior distribution by a set of samples or particles, drawn from this distribution. Each particle is a concrete hypothesis of the true system state. The set of particles approximates the posterior distribution. PF is best described as a nonparametric Bayes filter because it develops the belief using a finite number of samples. 

Hypotheses of the system state at time $t$ is represented by a set $P_t$ of $N_P$ weighted random particles: 
\begin{equation} 
P_t =  \{  \langle X_t^{\nint{i}}, w_t^{\nint{i}} \rangle \ | \ i=1,...,N_P \}
\label{eq:system_state_particle}
\end{equation}

where $X_t^{\nint{i}}$ is the state vector of the \textit{i}-th particle and $w_t^{\nint{i}}$ is the corresponding weight. Weights are non-zero, and sum over all weights is 1. The core idea of the PF is to update and maintain this set of particles given model outputs and observations. A PF recursively estimates the particle set $P_t$ based on the estimate $P_{t-1}$ at the previous time step, and the observation. The PF algorithm can be briefly described in three steps:
\begin{enumerate}
	
	\item \textbf{Predict:} Generate the next set of particles $\hat{P}_t$ from the previous set $P_{t-1}$. This represents the \textit{prior} distribution to describe how the system state evolves. 
	\item \textbf{Importance Weighting:} Compute the importance weight $w_t^{\nint{i}}$ for each particle in $P_t$. This is equivalent to the `Update' step in Kalman Filter, and will give us the \textit{posterior} distribution 
	\item \textbf{Resampling:} This step has no analogous step in Kalman Filter and its extensions. The resampling step creates a new set of particles from the current set. The likelihood to draw a particle is proportional to its weight. We adopt Sample Importance Resampling (SIR), a popular bootstrap systematic resampling in the PF literature \citep{wang_data_2015, carrassi_data_2018}. SIR has been developed to deal with \textit{particle deprivation}, which is the problem when particles converge to a single particle after several iterations due to one particle outperforming all others \citep{kong_sequential_1994}. This problem significantly reduces the area of state space covered by the particles in later iterations. 
\end{enumerate}
Since resampling will generate particles using the existing pool of particles, it will not be able to produce particles where the prediction accuracy is better than the existing particle pool. This means that in \textit{classical} PF, the model parameter $S_t$ of both BusSim-deterministic and BusSim-stochastic will be unchanged over time. Because the  parameters change over time, we need to dynamically optimise $S_t$. This problem is solved in this paper by a simple and generic solution. We improve the quality of the particles by diversification similar to \citep{vadakkepat_improved_2006}, in a process also known as roughening, jittering, and diffusing \citep{pantrigo_combining_2005}. This is achieved by adding a random Gaussian white noise $\sigma$ with mean 0 and a predefined standard deviation, not to the whole state vector $X_t$, but to the model parameter $S_t$, to increase the probability of having particles that represent the current state of the underlying model. 

The PF is applied to BusSim-deterministic and BusSim-stochastic using up-to-date data from the synthetic `real-time' GPS data. 

\section{Numerical experiment}
\label{s:experiments}
\subsection{Experiment set up}

To generate the synthetic `historical' and `real-time' GPS data used in BusSim-deterministic and BusSim-stochastic, we predetermine a set of model parameters to generate realistic GPS data. Table \ref{tab:param} lists the fixed parameters being used in this experiment. 

\begin{table}[htbp]
	\centering
	\caption{Fixed parameters in BusSim-truth}
	\begin{tabular}{rll}
		\multicolumn{1}{l}{\textbf{Class}} & \textbf{Parameter} & \textbf{Value} \\
		\midrule
		\multicolumn{1}{l}{Bus} & FleetSize  & Unique ID of the bus agent \\
		& Acceleration & 3 $m/s^2$ \\
		& [$\theta_1,\theta_2,\theta_3$] & [3,1,0.85] $s$ \\
		\midrule
		\multicolumn{1}{l}{BusStop} & Number of Stops & 20 \\
		& Length between stops & 2000m \\
		& GeoFence & 50m \\
		\bottomrule
	\end{tabular}%
	\label{tab:param}%
\end{table}%

Second, the dynamic parameter set $S_t = [Arr_m^t \ Dep_m^t \ V^t]$ (Equation \ref{eq:S}) are time-varying and therefore, randomly generated using fixed rules. We first generate an initial arrival rate $Arr_m^0$ at stop $m$ at time 0 by a random generation from an uniform distribution between the minimum and maximum passenger arrival rate [$minDemand, maxDemand$]. 

\begin{equation}
Arr_m = \mathcal{U} (minDemand, maxDemand) \quad m = 1,...,M
\label{eq:arrival_rate}
\end{equation}

The departure rate is also generated from an uniform distribution, but also ordered non-decreasingly to represent the fact that more passengers alight at the end of the route than at the beginning. The departure rate at the last stop (stop M) is set as 1 to let every remaining passengers to alight the bus at the last stop. 
\begin{equation}
Dep_m =  ordered \ (\mathcal{U} (0.05, 0.5)), \quad Dep_M = 1 \quad \& \quad m = 1,...,M
\label{eq:departure_rate}
\end{equation}

\subsection{The stochastic and dynamic nature of BusSim-truth}

\begin{figure}[h]
	\centering
	\includegraphics[width=1\textwidth]{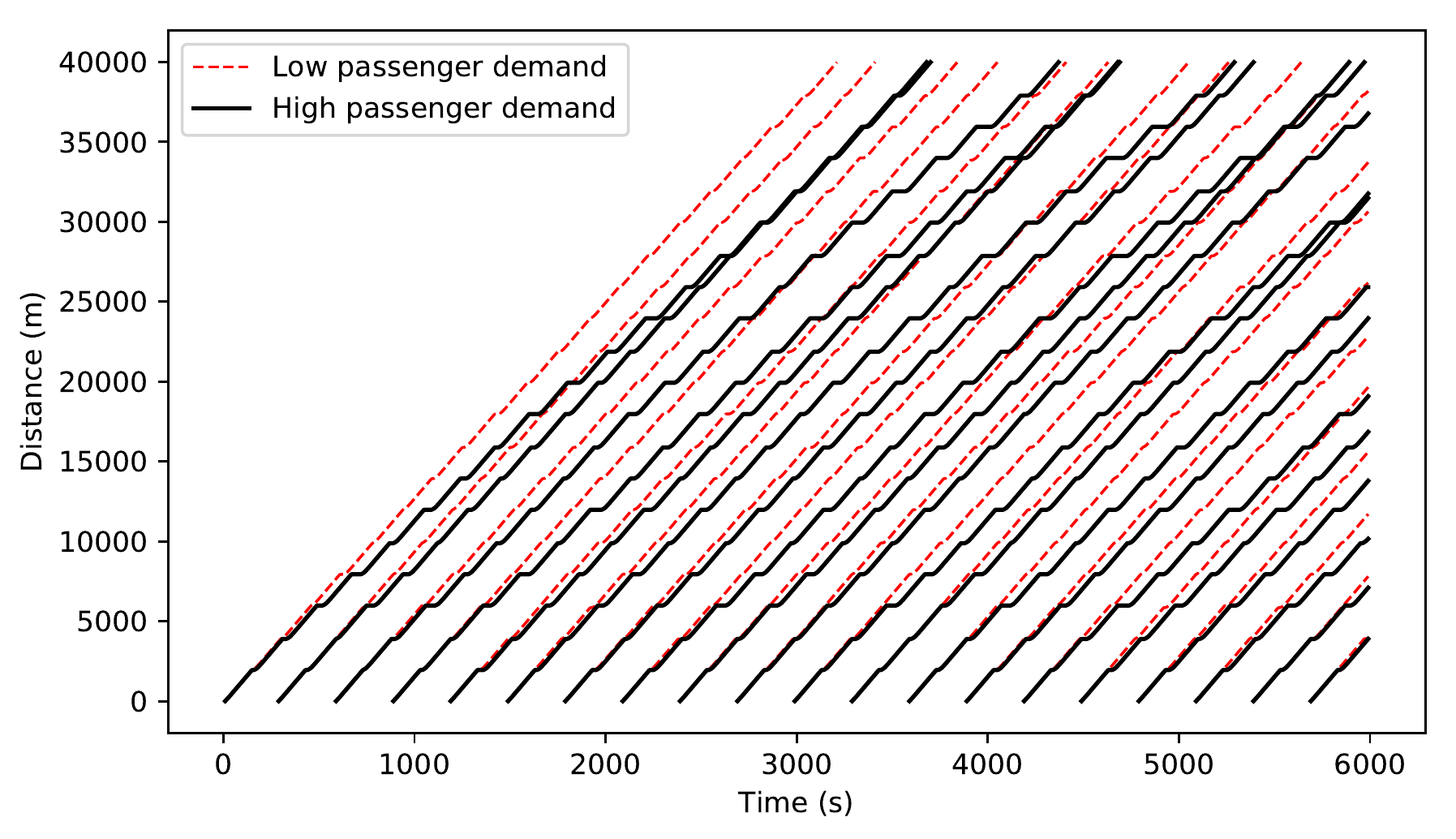}
	\caption{Synthetic bus GPS trajectory at low and high passenger demand. Red, dashed lines are bus trajectories when $maxDemand$ equals 0.5, while black, solid lines are bus trajectories when $maxDemand$ equals 2.}
	\label{fig:2stochastic}
\end{figure}

\begin{figure}[h!]
	\centering
	\includegraphics[width=1\textwidth]{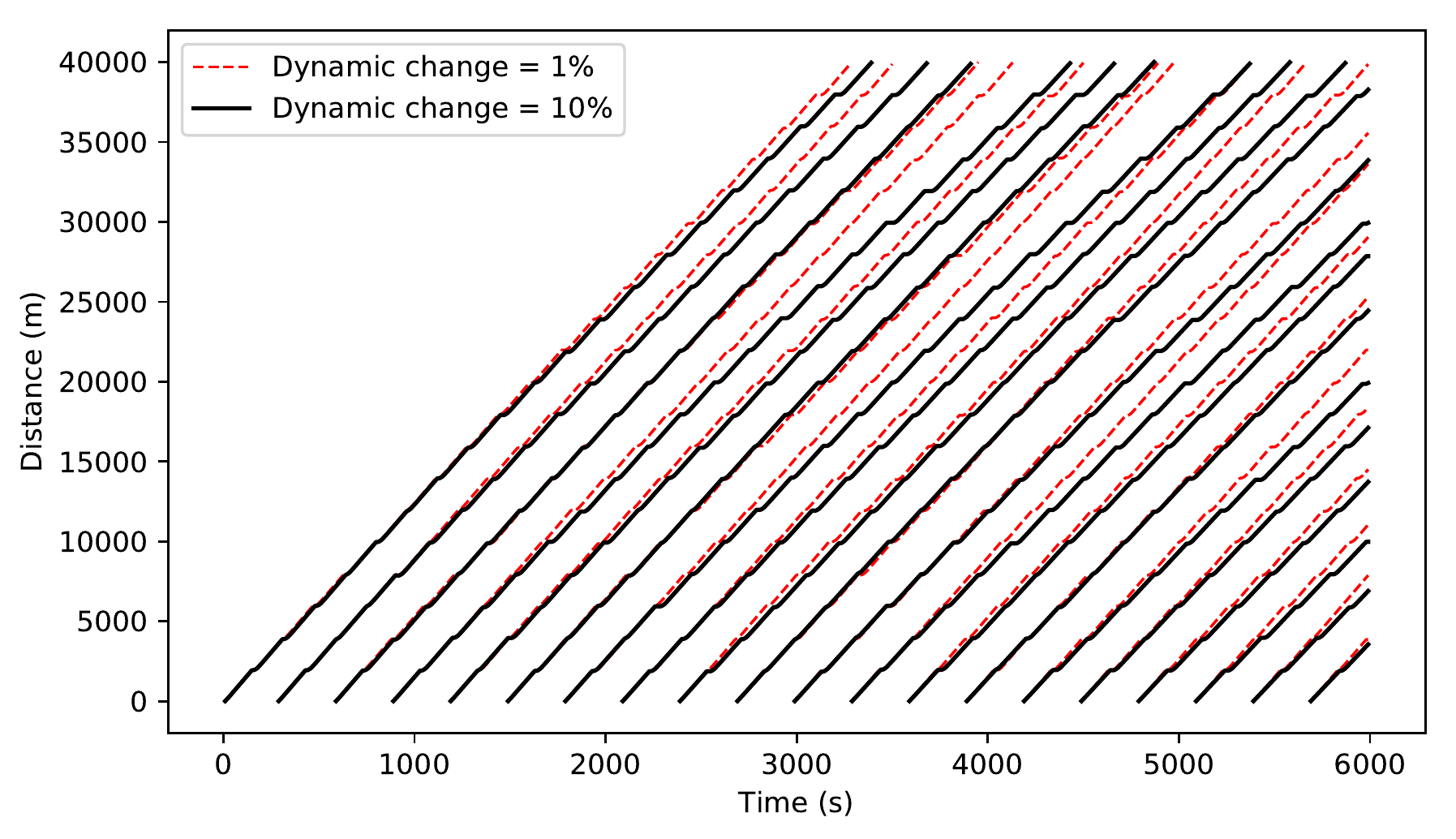}
	\caption{Synthetic bus GPS trajectory with two different value of $\xi$.}
	\label{fig:2dynamic}
\end{figure}

Figures  \ref{fig:2stochastic} and \ref{fig:2dynamic} provide a simple verification that demonstrates BusSim-truth generates realistic synthetic GPS data under different sets of parameters. Other variables have also been verified and will be used in the sensitivity analysis. This section outlines how this validation was achieved.

We aim to control the stochastic and dynamic level in BusSim-truth using only a single parameter for each. Equation \ref{eq:arrival_rate} controls the level of stochasticity in BusSim-truth. For instance, a pair of values [$minDemand, maxDemand$]=[0.5,1] means 0.5 to 1 passenger arriving at the bus stop each minute. By fixing the $minDemand$ to be a small number (e.g. equals 0.5), we can control the stochasticity of BusSim-truth by a single parameter $maxDemand$, with a larger $maxDemand$ meaning more stochasticity and vice versa. We control the level of dynamicity in BusSim-truth by a dynamic change rate parameter $\xi$, which gradually changes the arrival rate and surrounding traffic speed over the simulation period. 

To implement an inner verification of the BusSim-Truth model and to investigate the impacts of the \textit{stochastic} and \textit{dynamic} natures of the system under study, we evaluate the outputs from BusSim-truth under different values of stochasticity and dynamicity. Figure \ref{fig:2stochastic} gives an insight into the differences in bus trajectories when $maxDemand$ equals 0.5 and 2. Note that when $maxDemand$ equals 0.5, BusSim-truth reduces to a deterministic model (similar to BusSim-deterministic) because $maxDemand$ would then be equal to $minDemand$. 

Each line in Figure \ref{fig:2stochastic} shows the GPS trajectory of bus location, as generated by BusSim-truth. The solid lines show the trajectory of buses at high and stochastic demand ($maxDemand$ equals 3), whereas the dashed lines are for low and deterministic demand ($maxDemand$ equals $minDemand$). The trajectories in Figure \ref{fig:2stochastic} show that as the $maxDemand$ increases, there are more delays for each individual buses and less likely that buses are able to keep stable headway from each other. 

The dynamic nature of BusSim-truth is illustrated in 
Figure \ref{fig:2dynamic} when the dynamic change rate parameter $\xi$ is equal to 1\% and 10\%. Because the arrival rate and traffic speed gradually change, there is little change in the bus trajectories of BusSim-truth with $\xi$ equals 1\% and 10\%. As time passes, there are more delays for BusSim-truth with $\xi$ equals 10\%. This is because there are more passengers (higher arrival rate) and the buses are also travelling slower (lower traffic speed).

\subsection{Scenario 1: no calibration (benchmark)} 

\begin{figure}[h]
	\centering
	\includegraphics[width=1\textwidth]{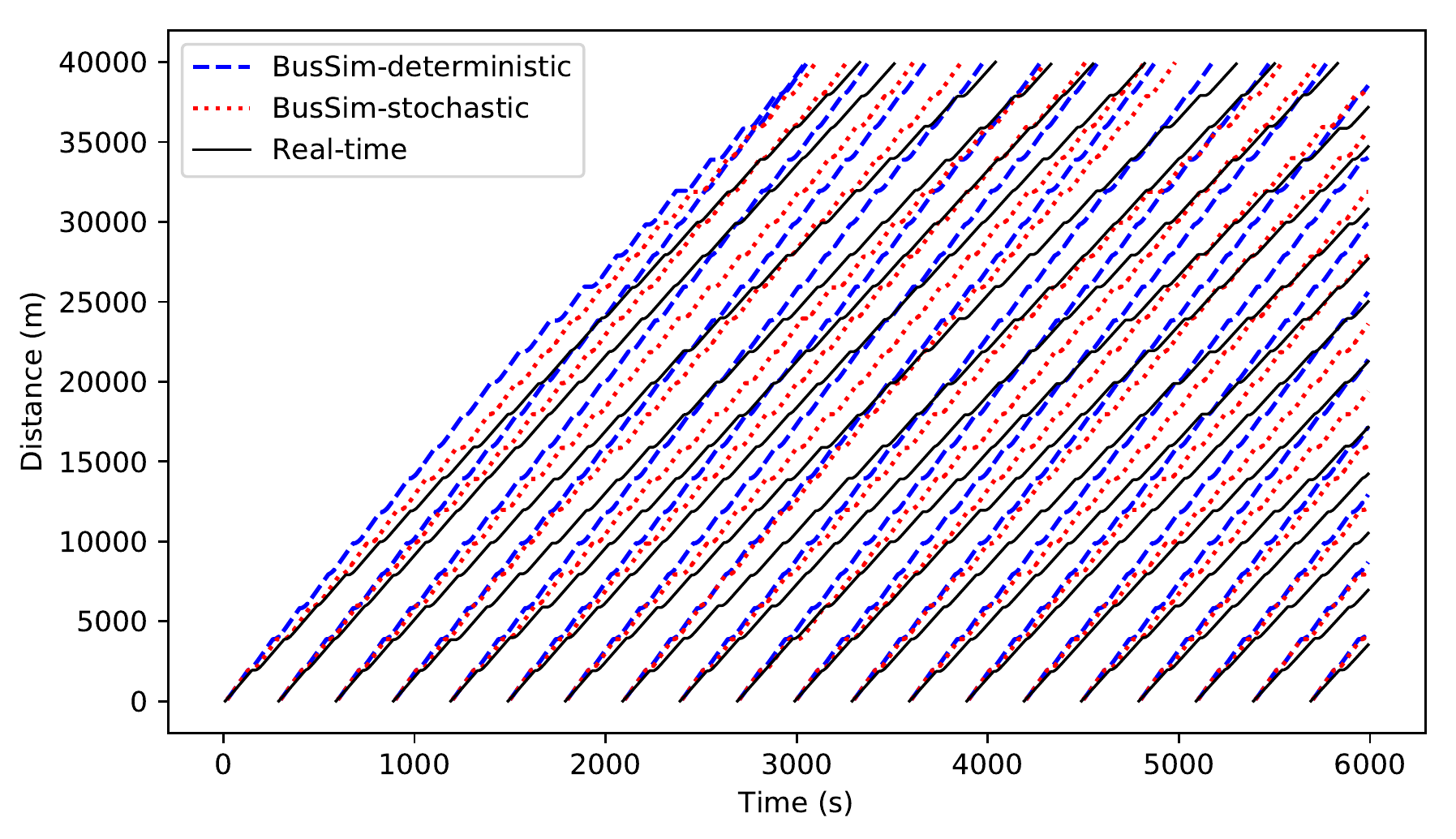}
	\caption{Prediction results from Scenario 1: no calibration}
	\label{fig:do_nothing}
\end{figure}

This scenario aims to evaluate the prediction results from BusSim-deterministic and BusSim-stochastic \textit{without} calibrating their parameters or performing data assimilation. This is necessary so that later we can evidence the additional predictive performance of the model after calibration \textit{and} data assimilation. The two models are implemented using random parameters generated from Equation \ref{eq:arrival_rate} and \ref{eq:departure_rate}. The outputs from these models are bus locations at each time step $t$, which can be compiled to space-time trajectories and compared to the synthetic `real-time' bus trajectories, as illustrated in Figure \ref{fig:do_nothing}. 

Figure \ref{fig:do_nothing} shows one particular case where $maxDemand$ equals 2, and $\xi$ equals 7\%, as an example of the prediction results. Both models poorly predict the trajectories of the `real' buses. This is expected because the models do not have the optimal parameters to capture the bus route operations. These models are therefore not useful for real-time prediction without parameter calibration or data assimilation. 

\subsection{Scenario 2: Parameter calibration}

\begin{figure}[htb]
	\centering
	\includegraphics[width=1\textwidth]{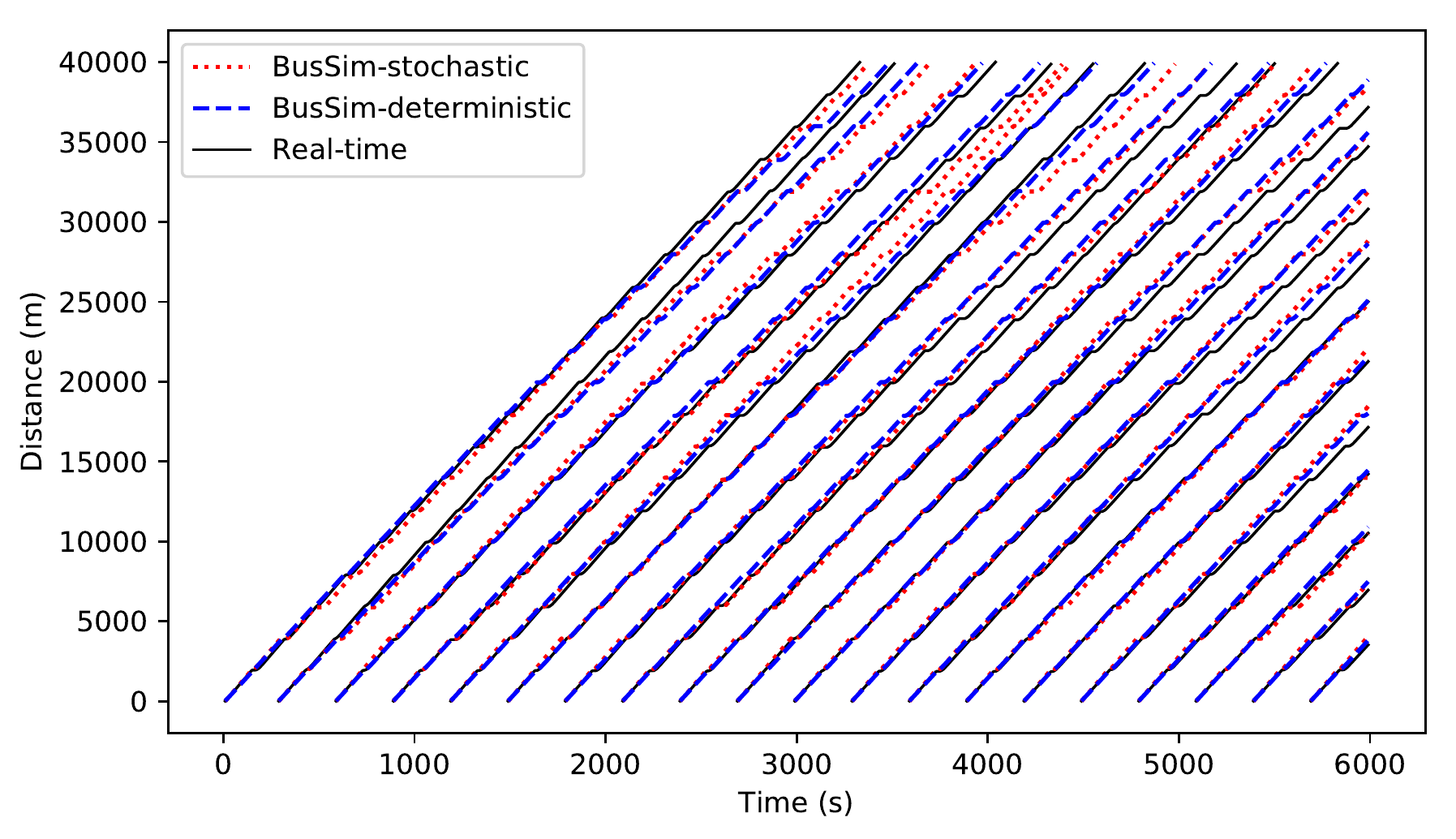}
	\caption{Prediction results from Scenario 2: Parameter calibration}
	\label{fig:calibration}
\end{figure}

In this scenario, BusSim-deterministic and BusSim-stochastic are calibrated using the Cross-Entropy Method, as described in Section \ref{s:calibration}. The two calibrated models are used to predict the bus locations at each time step $t$, which can be compiled to trajectories. Figure \ref{fig:calibration} shows an example of the comparison between the predictions from BusSim-deterministic and BusSim-stochastic versus the synthetic `real-time' GPS data, where the $maxDemand$ equals 2 and $\xi$ equals 7\%. 

Although they are improvements compared the un-calibrated versions -- Figure \ref{fig:calibration} shows that both models outperform the models in the Scenario 1 (no calibration [see Figure \ref{fig:do_nothing}]) -- the models can only predict well early in the simulation when there is little deviation in passenger arrival rate and surrounding traffic speed. There are large observable gaps between the predictions from BusSim-deterministic and BusSim-stochastic as buses reach the end of their routes. The two models were trained with synthetic `historical' data, but evaluated with `real-time' data. Recall that there are differences between the `historical' and `real-time' data due to the stochastic nature of the system under study (see Figure \ref{fig:historical_realtime}). Therefore a data assimilation procedure is required to prevent the errors gradually increasing throughout the simulation.

\subsection{Scenario 3: Applying a Particle Filter}

\begin{figure*}[htb]
	\centering
	\includegraphics[width=1\textwidth]{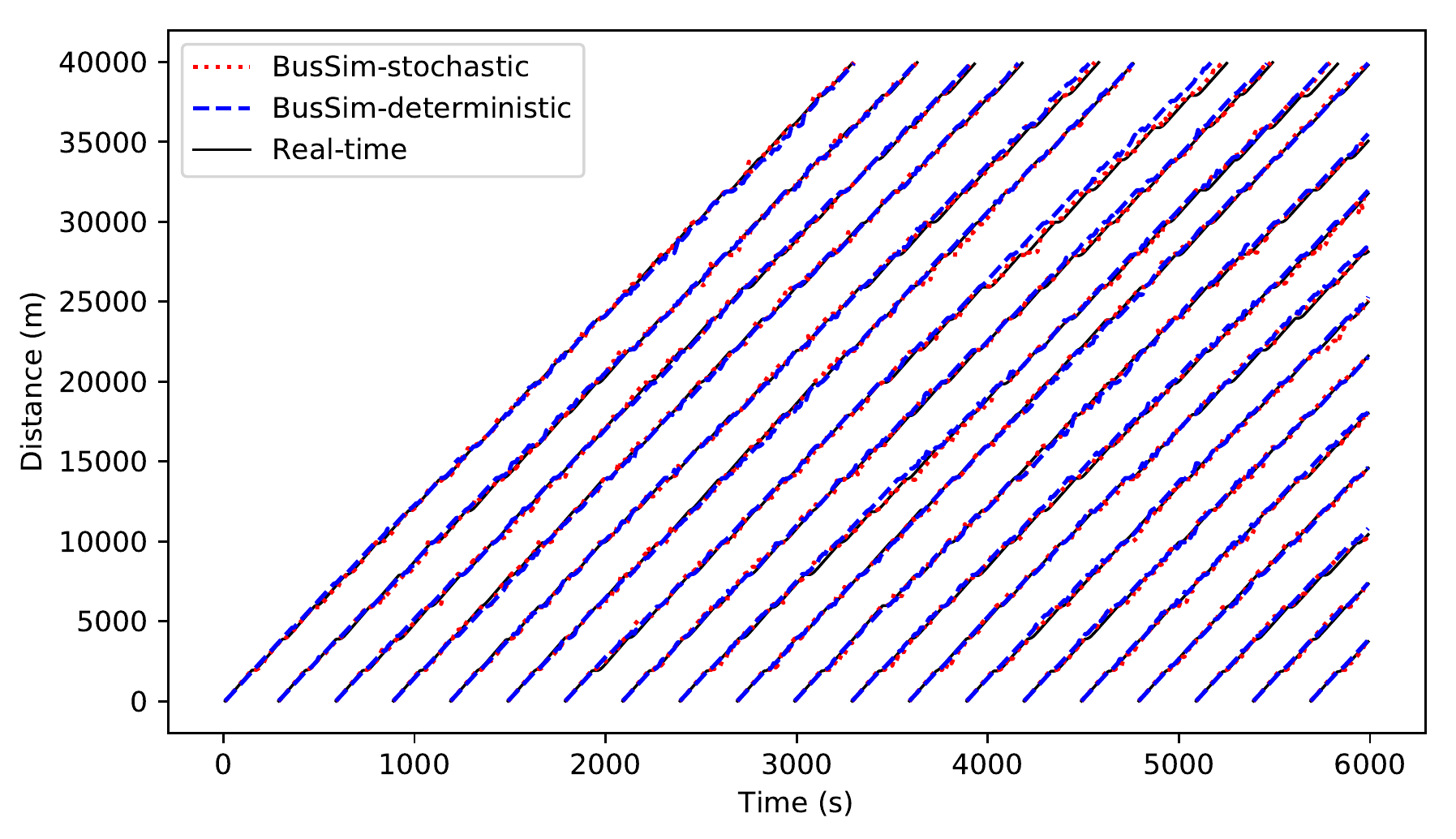}
	\caption{Prediction results from Scenario 3: Parameter calibration and Particle Filtering}
	\label{fig:calibration_plus_PF}
\end{figure*}

This section applies a PF to the calibrated BusSim-deterministic and BusSim-stochastic, as described in section \ref{s:PF}. At each time step $t$, the two models are only provided with the observation vector $O_t$, and then attempts to use $O_t$ to correct their prediction of future state vectors $X_t$ to $X_T$, where $T$ is the last time step. Figure~\ref{fig:calibration_plus_PF} illustrates the results after the models have been calibrated \textit{and} have `real-time' data incorporated (assimilated) into them during runtime.

The predicted bus trajectories in Figure~\ref{fig:calibration_plus_PF} fit much closer to the synthetic `real-time' data than the previous scenarios (Figure \ref{fig:do_nothing} and \ref{fig:calibration}). There are still observable gaps between the prediction and the synthetic `real-time' GPS data, because the underlying models do not know the underlying stochasticiy and dynamicity in the synthetic data, but the improvements (which will be quantified shortly) certainly \textit{appear} to be substantial.

\subsection{Sensitivity analysis}

In this section, we perform a sensitivity analysis to compare the prediction error in each scenario. The same experiments, as described in Scenario 1 to 3, are repeated at different values of $maxDemand$ and dynamic change rate $\xi$. To increase the robustness of the comparison, 10 replications have been made for each experiment, and the average Root Mean Squared Error (RMSE) values are reported. RMSE is calculated as the difference in prediction bus location and synthetic `real-time' bus location: 

\begin{equation}
RMSE = \sqrt{\frac{1}{T}\sum_{k=1}^{T}{\Big(\hat{y}_k -y_k\Big)^2}}
\end{equation}

Where $\hat{y}_k$ and $y_k$ is the bus location at time $k$ from the model prediction and synthetic `real-time' data, respectively. Table \ref{tab:sensitivity} compares the RMSE from each scenario. It is clear that the Scenario 3 (combination of parameter calibration and data assimilation) outperforms the other two Scenarios. 

\begin{table}[ht]
	\centering
	\caption{Sensitivity analysis of $maxDemand$ and dynamic change rate $\xi$}
	\begin{tabular}{ccccc}
		\toprule
		& \textbf{Values} & \textbf{Scenario 1} & \textbf{Scenario 2} & \textbf{Scenario 3} \\
		\midrule
		\multicolumn{1}{c}{\multirow{9}[2]{*}{maxDemand }} & 0.5   & 302   & 102   & 24 \\
		& 1     & 313   & 107   & 25 \\
		& 1.5   & 319   & 112   & 35 \\
		& 2     & 335   & 125   & 49 \\
		& 2.5   & 340   & 119   & 52 \\
		& 3     & 337   & 127   & 62 \\
		& 3.5   & 346   & 133   & 66 \\
		& 4     & 338   & 148   & 59 \\
		& 4.5   & 341   & 145   & 55 \\
		\midrule
		\multicolumn{1}{c}{\multirow{8}[2]{*}{Dynamic change rate}} & 0     & 197   & 75    & 41 \\
		& 2.5   & 203   & 77    & 44 \\
		& 5     & 208   & 82    & 40 \\
		& 7.5   & 211   & 89    & 39 \\
		& 10    & 218   & 90    & 49 \\
		& 12.5  & 220   & 93    & 47 \\
		& 15    & 232   & 97    & 45 \\
		& 17.5  & 235   & 102   & 49 \\
		\bottomrule
	\end{tabular}%
	\label{tab:sensitivity}%
\end{table}%

\section{Implications\label{s:implications}}

This paper presents an integrated framework to reduce uncertainty in ABMs when making predictions in real time, by combining parameter calibration and data assimilation. As discussed in Section \ref{s:Intro} and \ref{s:problem}, an `identical twin' approach has been adopted instead of real noisy data to facilitate an effective evaluation of the proposed methods against the synthetic `ground truth'. The numerical experiment shows that the framework yields more accurate predictions than (i) a benchmark scenario (without parameter calibration), and (ii) a scenario with parameter calibration but without data assimilation. 

In its current form, the framework can provide \textit{real time} bus locations and arrival times for passenger information systems. The forecasted bus location and arrival information provides key intelligence for waiting passengers \cite{fan2016waiting}. This is beneficial for all public transport passengers, but can be of particular benefit in countries, for example in the Global South \cite{kumar2017bus} where  there are frequent delays due to transport systems being complex, heterogeneous or heavily congested. The prediction of bus arrival times is also critical for real-time trip planners. These planning systems propose optimal alternative routes for passengers, or update information on a connecting service that may be unreachable due to delayed buses. 

Many advanced Intelligent Transport System applications heavily rely on predictions of bus location and arrival times, for  example  bus control studies such as \cite{daganzo2009headway}.  A model-based prediction of bus location and arrival time, such as the framework in this paper, would allow bus operators the ability to evaluate and update their transportation infrastructures in real time.
\section{Conclusion\label{s:conclusion}}

This paper proposes parameter calibration and data assimilation frameworks to enhance the prediction accuracy in agent-based models (ABM) when the system under study has a \textit{stochastic} and \textit{dynamic} nature. This is done in a 'identical twin' approach. We first develop a stochastic and dynamic ABM of bus route, referred to as \textit{BusSim-truth}. This model is employed to generate synthetic `historical' and `real-time' GPS data of bus locations. The `historical' data is used to train two simpler models of bus route, referred to as \textit{BusSim-deterministic} and \textit{BusSim-stochastic}, and evaluate against the 'real-time' data. 

Similar to the practice, when any simulation model is a simplification of the reality, BusSim-deterministic and BusSim-stochastic are simpler than BusSim-truth, and thus may not be able to produce a prediction similar to the synthetic `real-time' GPS data under limited data. We propose a solution for this issue by parameter calibration using Cross-Entropy Method (Scenario 2), by a combination of parameter calibration and Particle Filtering (Scenario 3), and show that they outperform the no calibration scenario (Scenario 1) and only Particle Filtering scenario (Scenario 4), at various levels of uncertainty. 

This paper shows the need for parameter calibration and data assimilation, and particularly the combination of them, to improve the accuracy of model-based prediction using ABMs in real time. Future research direction includes fitting the proposed framework with real data instead of synthetic data. 

\section*{Data Availability} This paper does not use any real data. Synthetic data has been generated from one of its models (BusSim-truth model). The source code for all the models, and the used synthetic data are available from \url{https://github.com/leminhkieu/Bus-Simulation-model}. 

\section*{Competing interests} We declare we have no competing interests

\section*{Acknowledgements} This project has received funding from the
European Research Council (ERC) under the European Union Horizon 2020 research and innovation programme (grant agreement No. 757455), a UK Economic and Social Research Council (ESRC) Future Research Leaders grant (ES/L009900/1) and a ESRC/Alan Turing Joint Fellowship
(ES/R007918/1).
\section*{Appendix A: The BusSim model \label{appendix:BusSim}}

Figure 2 illustrates the workflow for BusSim-truth. At each current time step $t$, each Bus agent checks whether the next time step would be larger than the vehicle's scheduled dispatch time $\delta_j$. If $t>\delta_j$, we then check whether the bus is on the road (Status equals $MOVING$), or at a stop for passenger dwelling (Status equals $DWELLING$), or has finished its service (Status equals $FINISHED$), otherwise the bus remains $IDLE$. 

If the status is $MOVING$, we first check whether the bus is at a bus stop, by comparing the $GeoFence$ area of each bus stop agent with the bus' location. If the bus is not approaching a bus stop, its current speed $v_j$ will be compared with the surrounding traffic speed $V$. If $v_j<V$, we assume that the bus will speed up with an acceleration rate $a_j$, thus we have: 
\begin{equation}
v_j^{t} = v_j^{t-dt} + a_j \cdot dt
\end{equation}

Therefore for the next time step, the bus will cover a distance of: 
\begin{equation}
S_j^t = S_j^{t-dt} + v_j^t \cdot dt
\end{equation}

If the speed already matches the traffic speed $V$, the bus will maintain the same speed. Or else if the bus is approaching a bus stop, the system will first check if the stop is the last stop. If it is the last stop, then the bus' status will be changed to $FINISHED$ and bus speed is changed to zero. If it is not the last stop, the system will change the status of agent Bus $j$ to $DWELLING$ and its speed to zero. The number of boarding and alighting passengers from the bus $j$, and the time that it will leave the stop are estimated as follows.  

The number of boarding passenger is proportional to the time gap between the current time (when Bus $j$ approaches the bus stop $m$) and the last time any bus visits the bus stop $m$:     
\begin{equation}
B_{j,m} = \nint{Po(Arr_m \cdot (t^a_{j+1,m}-t^a_{j,m}) } \quad | \quad B_{j,m}\in\mathbb{N}
\label{eq:Boarding_est}
\end{equation}

Equation \ref{eq:Boarding_est} shows that the number of boarding passengers is estimated using a stochastic Poisson process. A Poisson process is widely adopted in literature to estimate the count of passengers waiting at a public transport stop \citep{toledo2010mesoscopic,cats2010mesoscopic}. Extensions of this stochastic process have been introduced, such as non-homogeneous Poisson process \citep{kieu2018stochastic}, where the arrival rate is time-dependent, but for simplicity we adopt a homogeneous Poisson process for this paper. Equation \ref{eq:Boarding_est} makes the BusSim-truth model stochastic, because there is randomness in the way the Poisson process generates a number. For more details on the number generation process using stochastic Poisson process (e.g. thinning algorithm), interested readers may refer to \citep{lewis1979simulation}. The number of boarding passengers is also limited by the available capacity of the bus:
\begin{equation}
B_{j,m} = \text{max} \big( B_{j,m}, C - Occ_m )   \big)
\label{eq:Boarding_limit}
\end{equation}

The number of alighting passengers is proportional to the number of passenger on board (bus occupancy) and the departure rate at the stop $m$.  For simplicity, we assume that $A_{j,m}$ is the product between the departure rate from bus stop $m$ and the current bus occupancy (the number of passenger on board leaving the last stop): 
\begin{equation}
A_{j,m} = \nint{Dep_m \cdot Occ_{j,m-1}} \quad | \quad A_{j,m}\in\mathbb{N}
\end{equation}

To estimate the amount of time that bus will have to stay at the bus stop $m$ for passenger boarding and alighting, a.k.a. \textit{dwell time} $D_{j,m}$, we adopt the approach in \citep{bertini2004modeling} and the Transit Capacity and Quality of Service Manual (TCQSM) \citep{kfh2013transit}:
\begin{equation}
D_{j,m} = \theta_1 + \theta_2 \times B_{j,m} + \theta_3 \times A_{j,m} 
\label{eq:dwell_time}
\end{equation}
The parameter set [$\theta_1,\theta_2,\theta_3$] represents the time spent for passenger boarding, alighting, and a fixed value for vehicle stopping and starting, respectively. Equation \ref{eq:dwell_time} is the formulation for a single-door bus system, where boarding and alighting occurs sequentially. 

The departure time of bus $j$ from stop $m$ is calculated from the arrival time $t^a_{j,m}$ plus the time spent at stops for passenger boarding and alighting, or in other words the dwell time $D_m$:
\begin{equation}
t^d_{j,m} = t^a_{j,m} + D_{j,m}
\end{equation}
In BusSim, the bus $j$ is only allowed to leave the bus $m$ at time $t^d_{j,m}$, so this is also called the $Leave\_stop\_time$, as can be seen in the Figure 2. 

If the status of bus $j$ is $DWELLING$, it is at a stop for passenger boarding and alighting. We then check if the next time step would be larger or equal to the leave stop time $t^d_{j,m}$. If it would, then the bus would start accelerate to leave the stop, otherwise it would stay for at least another time interval. Finally, if the status of the bus is $FINISHED$, then we would do nothing. The modelling process then moves to the next Bus agent until the last Bus, then the whole model moves to the next time step until the last time step. 

BusSim-truth also assumes that parameters dynamically change over time by introducing an additional parameter $\xi$ to represent the change in passenger demand or surrounding traffic speed. For simplicity, we assume that a single, deterministic parameter $\xi$ can model these dynamic changes. In practice, it is possible, and more desirable, to use a time-dependent value of $\xi$ such that dynamic change is better captured, and multiple $\xi$ to model different changes. $\xi>0$ represents an increase in passenger demand and traffic speed, and $\xi<0$ represents otherwise. In this paper, the change in passenger demand or traffic speed is modelled as: 
\begin{align}
V = V \cdot \big( 1 - \frac{t}{T} \cdot \frac{100}{\xi} \big) \\
Arr_m = Arr_m \cdot (1 - \frac{t}{T} \cdot \frac{100}{\xi}
\label{eq:dynamic_bussim}
\end{align}
A positive value of $\xi$ in Equation \ref{eq:dynamic_bussim} gradually reduces the surrounding traffic speed $V$ and increases the arrival rate $Arr_m$, which would lead to more bus delays and congestion. 
\section*{Appendix B: Cross Entropy Method for Parameter Calibration}

This Appendix describes the pseudocode for the Cross Entropy Method for Normal distribution \citep{rubinstein1999cross}. 

\begin{algorithm} [H]	
	\SetAlgoLined
	Set $p=(\mu_1,\sigma_1,\mu_2,\sigma_2,...,\mu_K,\sigma_K)$ \quad \%Initial distribution parameters 
	
	Set $M$ \quad     \%Number of stops
	
	Set $T$  \quad  \% Maximum iteration number
	
	Set $I$  \quad  \% Maximum iteration number 
	
	Set $\rho$  \quad \% Set selection ratio
	
	\For {$t$ from 1 to $T$} { \  \%Main CEM loop
		
		\For {$i$ from 1 to $I$} {
			
			Draw $y^{(i)}$ from $\mathcal{N}(\mu,\sigma)$  \quad  \%Draw $I$ samples
			
			Compute $f^i:=f(y^{(i)}$}
		
		Sort $f^i$-values \quad  \%Order by decreasing magnitude
		
		$\gamma \leftarrow f_{\rho.I}$ \quad  \%Set threshold 
		
		$L_\gamma \leftarrow \{ y^{(i)} | f(y^{(i)}) \leq \gamma$  \quad  \%Collect elite samples 
		
		$\mu'_j = \frac{1}{L_\gamma} \sum_{i=1}^{L_\gamma} \mu_{i,j}$  \quad  \%Update $\mu$
		
		$\sigma'_j = \frac{1}{L_\gamma} \sum_{i=1}^{L_\gamma} \sigma_{i,j}$   \quad  \%Update $\sigma$
		
		$\mu_j \leftarrow \alpha \mu'_j + (1 - \alpha)  \mu_j$ \quad \%Update with step size $\alpha$
		
		$\sigma_j \leftarrow \alpha \sigma'_j + (1 - \alpha)  \sigma_j$ \quad \%Update with step size $\alpha$
	}
	\caption{Cross-Entropy Method for Normal distribution}
	\label{algo:CEM}
\end{algorithm}

\bibliographystyle{plain} 
\bibliography{2018-pf-bussim}

\end{document}